\documentclass[sigconf]{acmart}

\usepackage{multirow}
\setcopyright{none}
\renewcommand\footnotetextcopyrightpermission[1]{}
\usepackage{subfigure}

\newcommand{\rem}[1]{}
\usepackage{subfigure}
\begin{document}
\fancyhead{}
\title{Algorithmic Copywriting: Automated Generation of Health-Related Advertisements to Improve their Performance}

\author{Brit Youngmann}
\affiliation{\institution{Microsoft Research\\ Herzliya, Israel
}}
\email{t-bryoun@microsoft.com}

\author{Elad Yom-Tov}
\affiliation{\institution{Microsoft Research\\ Herzliya, Israel
}}
\email{eladyt@microsoft.com}

\author{Ran	Gilad-Bachrach}
\affiliation{\institution{Microsoft Research\\ Herzliya, Israel
}}
\email{rani.gb@gmail.com}

\author{Danny Karmon}
\affiliation{\institution{Microsoft Healthcare NExT\\ Herzliya, Israel
}}
\email{Danny.Karmon@microsoft.com}

\begin{abstract}

Search advertising, a popular method for online marketing, has been employed to improve health by eliciting positive behavioral change. However, writing effective advertisements requires expertise and experimentation, which may not be available to health authorities wishing to elicit such changes, especially when dealing with public health crises such as epidemic outbreaks.

Here we develop a framework, comprised of two neural networks models, that automatically generate ads. First, it employs a generator model, which create ads from web pages. It then employs a translation model, which transcribes ads to improve performance. 

We trained the networks using 114K health-related ads shown on Microsoft Advertising. We measure ads performance using the click-through rates (CTR). 

Our experiments show that the generated advertisements received approximately the same CTR as human-authored ads. The marginal contribution of the generator model was, on average, 28\% lower than that of human-authored ads, while the translator model received, on average, 32\% more clicks than human-authored ads. 
Our analysis shows that the translator model produces ads reflecting higher values of psychological attributes associated with a user action, including higher valance and arousal, and more calls-to-actions. In contrast, levels of these attributes in ads produced by the generator model are similar to those of human-authored ads.    

Our results demonstrate the ability to automatically generate useful advertisements for the health domain. We believe that our work offers health authorities an improved ability to nudge people towards healthier behaviors while saving the time and cost needed to build effective advertising campaigns.
\end{abstract}

\maketitle

\section{Introduction}

Spending on search advertising (also known as sponsored ads) in 2019 in the U.S. was valued at US\$$36.5$ billion~\cite{stat1} and US\$$109.9$ worldwide ~\cite{stat2}. The justification for these enormous amounts is the high efficiency of such ads, mostly due to the ability to tune the advertisements to explicit user intent, as specified in their queries, rather than on implicit information about their preferences \cite{garcia2011}. This, naturally, increases the likelihood of clicks and conversions (product purchases).

In search advertising, ads are shown on a Search Engine Results Page (SERP) whenever a user performs a search. Advertisers bid for specific keywords which they perceive as indicating an interest in their product. When these keywords are searched, their ads can be presented. As these ads are shown only for specific keywords, the displayed advertisement better matches the user's needs. The actual mechanism for matching ads to keywords are similar for most of the popular search engines (e.g., Google, Bing). Generally, search ads are targeted to match key search terms in the user's search query (namely the  keywords), which are provided by the advertiser, and additionally, advertisers may express preference for demographics (age, gender), location, and other user parameters. Advertisers compete with other advertisers who chose the same keywords, by submitting bids, representing monetary values for acquiring the ad slots associated with these keywords. The ads to be displayed are chosen by the advertising system using the bids, allowing advertisers to increase their exposure by bidding higher.

Various ad performance measures can be tracked \cite{garcia2011}, including the number of times an ad was shown (the number of impressions), the percentage of impressions which led to a click (the click-through rate, CTR), and the percentage of clicks which led to a purchase (the conversion rate). These performance measures are reported to the advertiser. The advertiser can optimize their campaign by modifying the ads, bids, or other related campaign parameters. Additionally, the advertiser can usually request that the advertising system optimize the campaign to one of the above-mentioned performance measures.

\begin{figure}[t]
\centering
\includegraphics[width=0.9\linewidth]{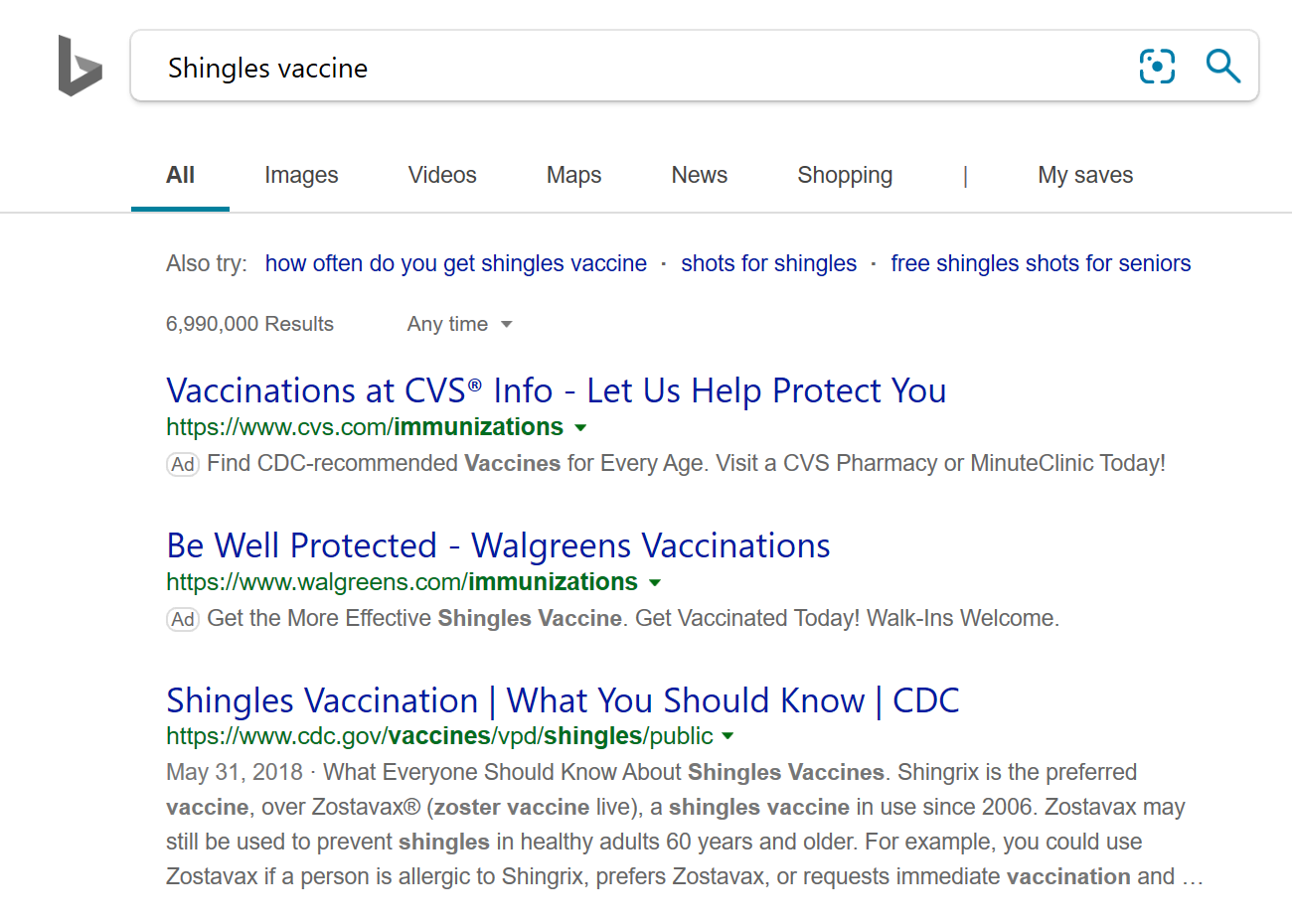}
\caption{Example Bing SERP with two search ads. The search query is ``Shingles vaccine" and the search ads appear as the two top results. }
\label{fig:serp}
\vspace{-2mm}
\end{figure}

Health authorities, pharmaceutical companies, and other stakeholders in the health domain have recognized that search advertising can assist not only in selling products, but (more importantly) in steering people towards healthier behaviors. Ads were shown to be effective in nudging people towards less harmful pro-anorexia websites~\cite{yomtov2018inducing}, to quit smoking~\cite{yomtov2016web} and towards increased physical activity and better food choices~\cite{yomtov2018effectiveness}. The conversion optimization mechanism of advertising engines was utilized (in conjunction with advertising campaigns) to improve HPV vaccination rates~\cite{mohanty2018} and to screen for cancer~\cite{arxiv}.

However, creating effective advertisements, as measured either using the above-mentioned performance measures or in eliciting behavioral change (the two are not synonymous~\cite{yomtov2018effectiveness}) requires expertise, experience, and testing. The first two are usually provided by advertising agencies, but these are generally proficient in selling products, not in promoting health outcomes. Testing ads is expensive and time-consuming. Thus, a health agency without the expertise or experience required to create an effective advertising campaign may not be able to do so, especially when the campaign needs to be quickly fielded in response to a public health crisis.

Here we offer a possible solution to this problem, based on recent advances in Natural Language Processing and on the vast repositories of ads and their performance, available to Internet advertising systems. We propose a framework that receives an (health-related) web page promoting healthy behaviour or selling some product or service, and generated from it an ad that maximizes a required performance measure. After discussing relevant prior work, we describe the building blocks of this framework. Specifically, we have divided the ad-generation task into two parts. In the first step we employ a generator pipeline, which receives as inputs URLs of web pages and generates ads from them. This pipeline includes a context extraction module, used to extract the relevant parts from the web pages, as well as out-of-the-box text summarization module, used to generating ads from the extracted text. In the second step, we employ a translator model, which receives as input the generated ads, and generate new ads that are expected to be optimized to achieve high performance. This pipeline includes ad normalization, preprocessing, and a sequence-to-sequence model.
%Both pipelines employ a deep neural network (DNN) sequence-to-sequence model.
For completeness of this work, we also report the results achieve by using solely the generator model, as well as the results achieved by using solely by the translator model (directly translating the original ads), allowing to estimate the marginal contribution of each of the pipelines.

We demonstrate the performance of the models and our insights into what they have learned to optimize the ads. Our experiments show that the advertisements generated by our full framework (i.e., the generator followed by the translator model) received approximately the same CTR as the human-authored ads (i.e., the original ads), implying that our framework can assist health authorities to automatically generate effective ads from scratch. The ads produced solely by the generator model received, on average, approximately 28\% fewer clicks than the human-authored ads. In comparison, the ads produced exclusively by the translator model received, on average, 32\% more clicks than the human-authored ads. This indicates that translating an ad to achieve better performance is an easier task than automatically generating an advertisement from a web page. Our analysis shows that the translator model produces ads reflecting higher values of psychological attributes associated with a user action, including higher valance and arousal, more calls to action, and the amplification of user desires, while the ads produced by the generator model behave similarly to the human-authored ads.

% We propose an algorithm that can take an (health-related) advertisement, for example, one created by a health agency, and rephrase it so as to maximize a required performance measure. After discussing relevant prior work we describe the building blocks of this algorithm, including ad normalization, preprocessing, and the training of a deep neural network (DNN) sequence-to-sequence model. We then show the performance of the algorithm and our insights into what the network learned in order to optimize the ads. Our experiments show that the proposed method increases the CTR on ads by $68\%$. Our analysis of the generated ads show that they reflect higher values of psychological attributes associated with user action, including higher valance and arousal, more calls to action, and the amplification of user desires. 

\section{Related Work}
\label{sec:related}
To the best of our knowledge, the proposed algorithms has no directly comparable past work. Instead, our work draws on past literature in several areas, including content extraction, abstractive summarization, machine translation, advertisement performance prediction, and work in psychology on the effectiveness of emotions in creating compelling advertisements. Here we review relevant work in these areas.

\subsection{Content Extraction from HTML documents}
% Parsing HTML documents is well studied task \cite{gupta2003dom} \cite{gupta2005automating} \cite{rahman2001content}. Exiting solutions enables users to quickly parse, extract, and manipulate data stored in HTML documents. Prominent practical tools include the Beautiful Soup \cite{BeautifulSoup} and Jsoup \cite{jsoup} libraries.

Web pages are often cluttered with distracting information around the body of an article, 
distracting users from the actual content they are interested in \cite{gottron2008content}. These information may include (pop-up or banner) advertisements, unnecessary images, or links scattered
around the text. Thus, automatic extraction of \emph{useful and relevant} content from web pages is a challenging task, which has extensively addressed in the literature \cite{sluban2013url} \cite{gupta2003dom} \cite{gupta2005automating} \cite{gottron2008content} \cite{peters2013content} \cite{song2015hybrid}. Automatic content extraction from web pages has many applications, ranging from enabling users to accessing the web pages more easily
over smart mobiles, to speech rendering for visually impaired, and text summarization \cite{gupta2005automating}.

% Gupta et al.~\cite{gupta2005automating} have proposed
% a content extraction technique that can remove clutter without destroying web page layout, creating a Document Object Model (DOM) tree, an approach
% also adopted by Chen et al. \cite{chen2003detecting}. 
% By parsing an HTML document into a DOM tree, one can extract information from large logical units, as well as manipulate smaller units such as specific links within the structure of the DOM tree. DOM trees are highly transformable, and thus can be easily used to reconstruct a complete web page.
We note, however, that our task is simpler, as we aim to extract only meaningful and bounded-size text from a web page, such as the main paragraph and title, with the goal of generating an advertisement describing the product or service it offers. Therefore, as we describe in Section \ref{subsec:generator}, in this work we adapted the simple, efficient \emph{Arc90 Readability algorithm} (\url{https://github.com/masukomi/arc90-readability}), that extracts the most important content from a web page. This commonly used algorithm was transformed into the Redability.com product, incorporated into Safari's Reader view, Flipboard, and Treesaver.

\subsection{Abstractive Summarization}
Text summarization is the process of automatically generating natural language summaries from an
input text, while retaining its main content. 
Generating short and informative summaries from large quantities of information, is a well-studied task \cite{dorr2003hedge} \cite{nallapati2017summarunner} \cite{nallapati2016abstractive} \cite{tas2007survey} \cite{gambhir2017recent}, applicable for multiple tasks such as creating news digests, search, and report generation.

There are two prominent types of summarization algorithms. First, extractive summarization algorithms form summaries by selecting sentences of the input text as the summery \cite{dorr2003hedge} \cite{nallapati2017summarunner}. Second,
abstractive summarization models build an internal semantic representation of the original text, then use it to create a summary that is close to what a human might express \cite{liu2019text} \cite{dong2019unified} \cite{paulus2017deep}. Abstraction may transform the extracted content by paraphrasing sentences of the source text. Such transformation, however, is computationally much more challenging than extractive summarization, involving both natural language processing and often a deep understanding of the domain of the original text.

Neural network models for abstractive summarization are typically based on the attentional encoder-decoder model for
machine translation \cite{paulus2017deep} \cite{liu2019text}.

State-of-the-art abstractive summarization techniques (e.g., \cite{liu2019text} \cite{dong2019unified}) employs Transformer based models that have shown advanced performance in many natural language generation and understanding tasks. In this current work, we employ the Bidirectional Encoder Representations from Transformers (BERT) \cite{devlin2018bert} for text embedding, the latest incarnation of pretrained language models which have recently advanced a wide range of natural language processing tasks \cite{young2018recent}. We have also used the model of \cite{liu2019text}, for summarizing the text extracted from web pages into ads.

% conceptualize the task as a sequence-to-sequence
% problem, where an encoder maps a sequence of
% tokens in the source text
% to a sequence of continuous representations, and a decoder then generates the target
% summary token-by-token, in an
% auto-regressive manner.

\subsection{Machine Translation}
Machine translation (MT) is a sub-field of computational linguistics that investigates the use of a machine to translate text from a source language to a target language, while keeping the
meaning and sense of the original text the same. 

The MT process can be simplified into three stages: the analysis of source-language text, the transformation from source-language text to target-language text
and the target-language generation. Work on MT can be divided into three main approaches: Rule-based MT \cite{nyberg1992kant} \cite{nirenburg1994machine} Statistical MT \cite{weaver1955translation} \cite{koehn2007moses} \cite{brown1988statistical} and Neural MT \cite{cho2014learning} \cite{papineni2002bleu}. 

In rule-based MT systems (e.g., \cite{forcada2011apertium}), a
translation knowledge base consists of dictionaries and grammar rules. A main drawback of this approach is the requirement of a very significant human effort to prepare the rules and linguistic resources, such as morphological analyzers, part-of-speech taggers and syntactic parsers. 

In statistical MT systems, the translations are generated on the basis of statistical models whose parameters are derived from the analysis of bilingual text corpora \cite{koehn2003statistical} \cite{chiang2005hierarchical}. Generally, the more human-translated text is available in a given language, the better the translation quality. A main drawback in statistical MT is that it depends upon huge amounts of parallel texts, and its inability to correct singleton errors made in the source language.

State-of-the-art MT systems use neural networks to predict the likelihood of a translation of a sequence of words \cite{kalchbrenner2013recurrent} \cite{sutskever2014sequence} \cite{cho2014properties}. As opposed to previous approaches, in neural MT, all parts of the translation model are trained jointly (end-to-end) to maximize performance, requiring minimal domain knowledge.
Such systems often use encoder-decoder architectures, encoding a source sentence into a fixed-length vector from which a decoder generates the
translation. 

In the current work we adopted a simple neural MT model to translate advertisements to ones attracting more users to click on the advertisements. We note that recent work has proposed an attentional mechanism to improve translation performance, by selectively focusing on
parts of the source sentence during translation. \cite{luong2015effective} \cite{bahdanau2014neural}. However,
as we show, even our simple neural MT model achieves significant improvement in terms of click-through rates (See Section Results). We note that a further improvement may be achieved by using an attentional-based model. Nonetheless, as the translation task is only one of the black-box components of our framework and not a part of our contributions, we leave this direction for future research.

%More recently, an attentional mechanism has been proposed to improve translation performance, by selectively focusing on
%parts of the source sentence during translation. $$\cite{luong2015effective, bahdanau2014neural}. 

\subsection{Click-through rate prediction}
As mentioned in the Introduction, popular search engines (such as Google and Bing) use keyword auctions to select advertisements to be shown in the display space allocated alongside search results. Auctions are most commonly based on a pay-per-click model where advertisers are charged only if their advertisements are clicked by users.
For such a mechanism to function efficiently, it is necessary for the search engine to
estimate the click-through rate (CTR) of ads for
a given search query, to determine the optimal allocation of
display space and payments \cite{graepel2010web}. As a consequence, the task of CTR prediction has been extensively studied \cite{guo2017deepfm}, \cite{juan2016field}, \cite{yan2014coupled}, \cite{zhou2018deep}, since it impacts user experience, profitability of advertising, and search engine revenue.

CTR prediction is based on a combination of campaign and advertiser attributes, temporal information, and, especially, the keywords used for advertising. While the former are dense, the latter are sparse, and hence need care in their representation.

Bengio et al. \cite{bengio2003neural} suggested learning a model based on a distributed representation for possible keywords, aiming to avoid the curse of dimensionality in language modeling. More recently, the authors of \cite{guo2017deepfm}, \cite{gai2017learning} proposed networks with one hidden layer, which first employ an embedding layer, then impose custom transformation functions for target fitting, aiming to capture the combination
relations among features. Other works (e.g., \cite{covington2016deep}, \cite{cheng2016wide}, \cite{shan2016deep})
suggested replacing
the transformation functions with a complex Multilayer Perceptron (MLP) network, which
greatly enhances the model capability. Generally, these methods follow a similar model structure with
combination of embedding layer (for learning the dense representation of sparse features) and MLP (for learning the combination
relations of features automatically). 

Predicting exact CTR values of ads is beyond the scope of the current work. As we explain in Section \ref{subsec:ranker}, here we focus on generating ads with higher CTR values than those of their human-authored ads, hence we use a simple ranking model for this task. However, our model is based on insights from the above-mentioned prior work. %An interesting direction for future work would be to predict exact CTR values of the generated ads.   

\subsection{The effect of emotion in advertising}

\begin{figure}[t]
\centering
\includegraphics[width=8cm]{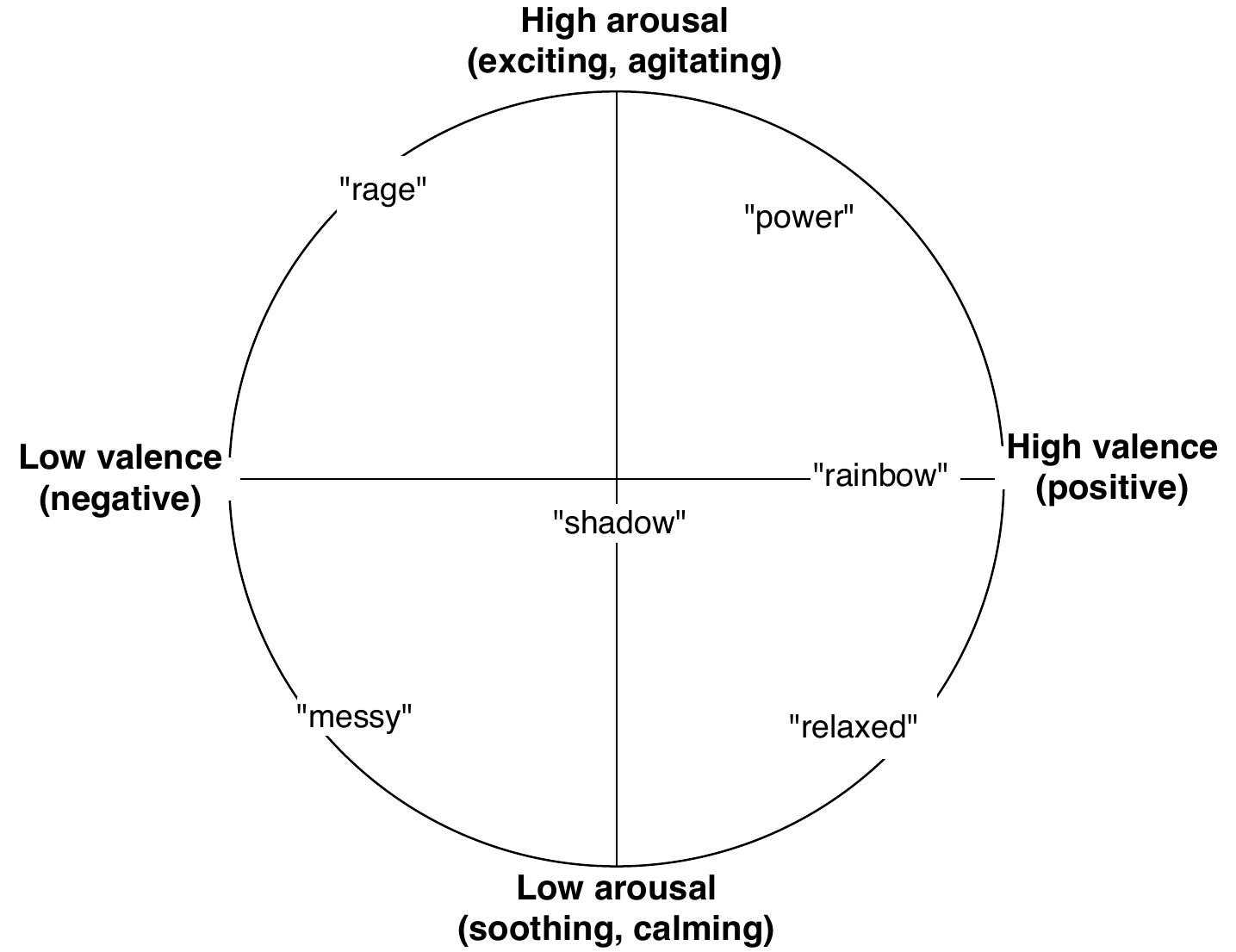}
\caption{Different English words mapped on the arousal-valence space. (Words were placed according to \cite{bradley1999affective}).  }
\label{fig:arousal_valence_circle}
\end{figure}

A widely-accepted framework proposes that affective (emotional) experiences can be characterized by two main axes: arousal and valence~\cite{kensinger2004remembering} \cite{lang1995effects} \cite{bradley1992remembering} \cite{conway1994formation} \cite{hamann2001cognitive}. The
dimension of valence ranges from highly
positive to highly negative, whereas the
dimension of arousal ranges from calming or
soothing on one end of the spectrum to exciting or agitating on the other. Figure~\ref{fig:arousal_valence_circle} shows an illustration of this, where different affective states are located in the space spanned by these axes~\cite{bradley1999affective}. 

Professionals of advertising aim to create ads which capture consumers' attention with the aim of increasing advertising
effectiveness. It has been shown that the inclusion of high arousal and valence sequences in ads increases user attention and interest~\cite{belanche2014influence} \cite{lang1995effects} \cite{lee2012effect}. 

Arousal is related to body activation
level in reaction to external stimuli~\cite{gould1992arousal}. Arousal has been related to simple
processes such as awareness and attention, but also to more complex tasks such as information retention~\cite{holbrook1982experiential}. Previous work suggests that arousal modulates ad effectiveness
and memory decoding~\cite{jeong2012there}.
However, highly arousing contexts can distract individuals from ad processing, making recall more difficult, thus reducing the ability to encode ad content~\cite{shapiro2002understanding}. 

In comparison to the considerable number of studies
investigating the effect of arousal on memory and attention,
relatively few studies have examined the effect of valence. The few studies which have examined its effect suggest
that valence is sufficient to
increase memory performance. Namely, non-arousing ads with
very positive or very negative valence are better
remembered than neutral ones~\cite{kensinger2004remembering} \cite{kensinger2003memory}. 

Ads which refer to the desires of users, especially those which take the form of special textual content, can affect CTRs in sponsored search \cite{wang2013psychological}. These desires can be mined from ad themselves using \emph{thought-based} effects and \emph{feeling-based} effects.
Thought-based effects are primarily related to win/loss analysis (e.g., trade-off between price and quality), while feeling-based effects are more subjective, and include, e.g., brand loyalty and luxury seeking.

%The authors of \cite{wang2013psychological} have demonstrated that the user psychological desire, especially in the form of special textual content, can affect the click-through rates in sponsored search. They have proposed to model user psychological desires, which are mined from ad texts, using \emph{thought-based} effects and \emph{feeling-based effects}. Thought-based effects are basically win/loss analysis (e.g., trade-off between price and quality), while feeling-based effects are more subjective (e.g., brand loyalty and luxury seeking).

In our analysis, we examine both the arousal/valence emotions in the human-authored ads compared to those in the generated ads, as well as the thought-based and feeling-based effects of the human-authored and generated ads.

\section{Methods}

We address the problem of automatically generating ads for the health domain. To this end, we have divided this task into two parts. In the first step we employ a generator model, which receives as inputs URLs of web pages and generates ads from them. Then, we employ a translator model, which receives as input the generated ads, and generate new ads that are expected to be optimized to achieve high performance.
% propose two pipelines: a translator and a generator. The translator pipeline receives as input ads in their original, whereas the generator pipeline receives as inputs URLs of web pages. Both pipelines generate ads that are expected to be optimized to achieve high performance.  
In the following, we describe each of these pipelines in detail.

We refer to the ads in our training set which were created by advertisers as ``human-authored'' ads and to ads generated by the proposed pipeline models as ``generated'' ads.

For completeness of this work, we also report the results achieve by the generator model without the last rephrasing step, as well as the results achieved by using solely by the translator model (directly translating the human-authored ads), allowing to estimate the marginal contribution of each of the models.

\subsection{Generator}
\label{subsec:generator}
Our proposed ads generator pipeline consists of two steps: content extraction, and ad generation. 

\subsubsection{Content Extraction}
Given an HTML document, our goal is to extract relevant and meaningful content from the document, contacting the most important information about the product/service to be sold.
This information may include details about the product/service (e.g., ``Get your birth control pills online"), special offers (e.g., discount, free/fast delivery), and a short explanation of the page content (e.g., ``Top 10 Foods to Help Manage Diabetes''). Another important restriction is that we want the extracted content to be short and concise. Unlike previous work that aim to separate the main content of a web page from the noise it contains (e.g., advertisements, links to other pages, etc.) \cite{gupta2005automating} \cite{gottron2008content} \cite{peters2013content} \cite{song2015hybrid}, our goal is to extract only a few main paragraphs from the web page.

\begin{figure}[H]
\begin{minipage}{0.5\textwidth}
\begin{figure}[H]
\includegraphics[width=1\textwidth]{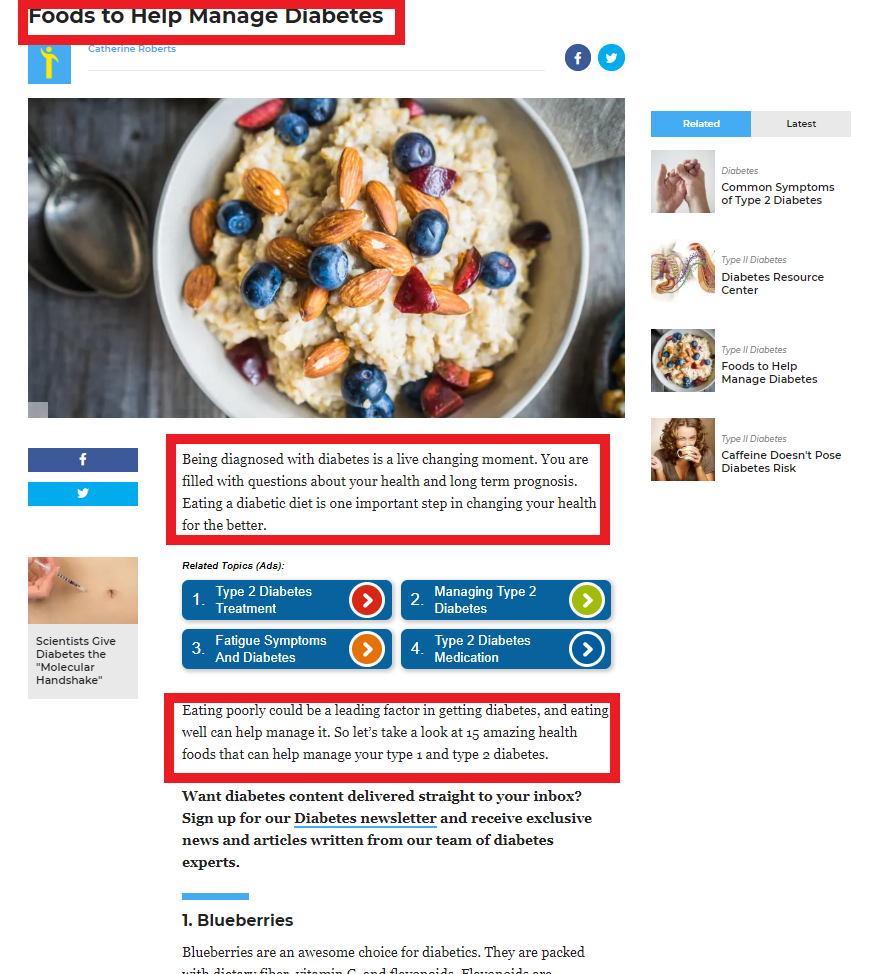}
%\caption{\label{fig:blue_rectangle} Rectangle}
\end{figure}
\end{minipage} \hfill
\begin{minipage}{0.45\textwidth}
Diabetes, What To Eat? - 15 Foods For Diabetes. Check Out These 15 Foods That Can Help You Control Blood Sugar.
\end{minipage}
\begin{minipage}{0.5\textwidth}
\begin{figure}[H]
\includegraphics[width=1\textwidth]{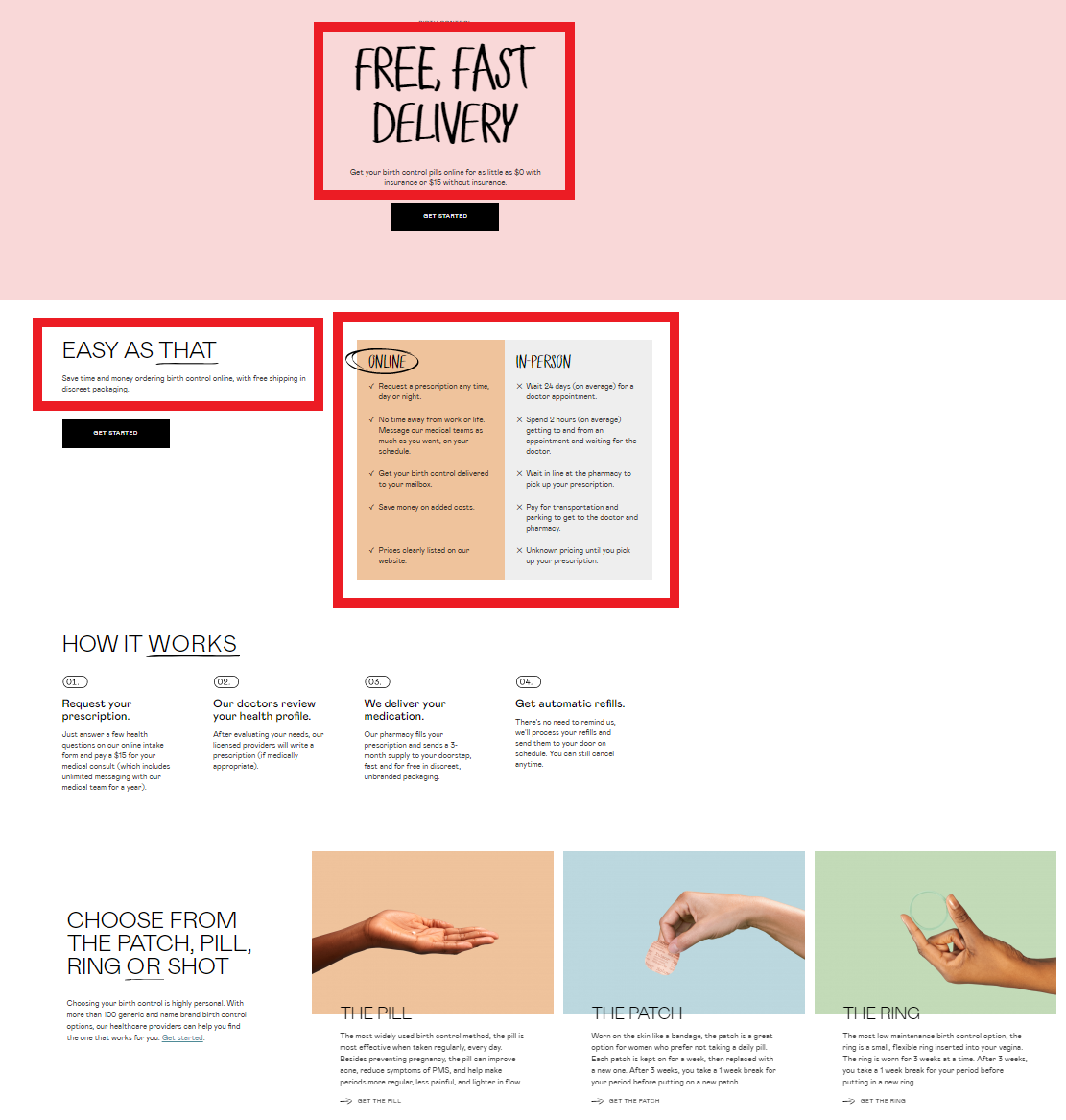}
\end{figure}
\end{minipage} \hfill
\begin{minipage}{0.45\textwidth}
Order Contraception Online - Safe, effective, convenient.	Learn about oral contraceptives from our licensed doctors \& free prescription.
\end{minipage}
\caption{Two examples of web pages along with their displayed ads. Marked in red are the relevant parts that were extracted from the pages.}
\label{fig:webpages}
\end{figure}

As the web pages have very different structures (e.g., as in Figure \ref{fig:webpages}), there is no precise ``one size fits all" algorithm that could
achieve this goal. Nonetheless, in this work, we have adopted a simple yet highly effective algorithm that extracts meaningful content from raw HTML documents. This algorithm is called the \emph{Arc90 Readability algorithm}. It was developed by the Arc90 Labs to make websites more comfortable to read (e.g. on mobile devices). For completeness of this work, we next briefly describe how this algorithm operates. 

The Arc90 Readability algorithm is based on two lists of HTML attributes (i.e., ids and classes names). One list contains attributes with a ``positive" meaning, while the second list contains attributes with a ``negative" meaning.  Intuitively, the algorithm operates as follows. For each paragraph (i.e., a p-tag), it
adds the parent of the paragraph to a list (if it is not already added), and
initialize the score of the parent with $0$ points. 
If the parent has a positive attribute, add points to the parent, otherwise, subtract points.
Lastly, get the top parents with maximal points and extract their textual content. 

Examples of attributes names we considered in our setting are depicted in Table \ref{tbl:pos_neg_attributes}. As for the scoring, for negative attributes, the algorithms subtract $2$ points, and for positive attributes, the algorithm adds $1$ point. Here we have extracted, from each page, its title, concatenated with the top $2$ parents with the maximal points. 

To illustrate, consider the two web pages presented in Figure \ref{fig:webpages}. The content extracted from them using the Arc90 Readability algorithm is marked in red.

\begin{table}[tb]
\centering
\begin{tabular}{ |p{30mm}|p{50mm}| } 
 \hline

`Negative" attributes & ``footer", ``copyright", ``location", ``style", ``comment", ``meta". \\
\hline
``Positive" attributes& ``content", ``text", ``title", ``body", ``article", ``page", ``description".\\
\hline

\end{tabular}
  \caption{Examples of ``negative" and ``positive" attributes used for the Arc90 Readability algorithm. }
  \label{tbl:pos_neg_attributes}
   \vspace{-8mm}
\end{table}

% \begin{figure*}[!t]
% \centering
% \begin{tabular}{p{56mm}p{26mm}} 
% \includegraphics[width=0.45\textwidth]{figures/webpage1_.PNG} & \begin{tabular}{@{}c@{}} Diabetes, What To Eat? - 15 Foods For Diabetes. Check Out \\ These 15 Foods That Can Help You Control Blood Sugar.\end{tabular}\\
% \includegraphics[width=0.45\textwidth]{figures/webpage2_.PNG}&
% \begin{tabular}{@{}c@{}}
%  Order Contraception Online - Safe, effective, convenient.	Learn about \\ oral contraceptives from our licensed doctors \& free prescription. \end{tabular}\\
% \end{tabular}

% \caption{Two examples of web pages along with their displayed ads. Marked in red are the relevant parts that were taken from the pages. }
% \label{fig:webpages}
% \vspace{-10px}
% \end{figure*}

\subsubsection{Candidate generation}
Given the extracted content from a raw HTML document, our next step is to generate an ad to this page.  
To this end, we have trained a state-of-the-art model for abstractive summarization, named the PreSumm model, presented in \cite{liu2019text}. The authors of \cite{liu2019text} have showcase how Bidirectional Encoder Representations from Transformers (BERT) \cite{devlin2018bert}, the latest incarnation of pretrained language models which have recently advanced a wide range of natural language processing tasks, can be usefully applied in text summarization. They have proposed a general framework for both extractive and abstractive models. In this work we are focusing on abstractive summarization rather on extractive one, as it allows the process of ``paraphrasing" a text than just simply summarizing it. Texts summarized using this technique tend to look more human-like and produce more condensed summaries.

\paragraph{Training data.}
The training data was constructed as follows: For every web pages, we have extracted its content and considered its corresponding (human-authored) ad that achieved the highest CTR. The reason for that is to allow the model to learn only how to generate ads which are expected to have high CTR values. We have also experimented with a model trained over all human-authored ads. However, as its results were inferior we omit them from presentation.  
For every (web-page, corresponding ad) pair, we generated an example in the training data. For training the model, we have used the default parameters of the abstractive summarization model as described in the implementation of \cite{liu2019text}.

% \subsubsection{Rephrasing}
% In the previous step we have generated ads from the web pages. We note that this task is much harder than the task of ads rephrasing, with the goal of CTR maximization. Therefore, to improve the generated ads, in the next step we used the trained translator model to rephrase the generated ads. 
% Here again, we consider the first ad generated by the translator as the candidate ad. 

% For completeness of this work, we also report the results achieve by the generator model without this last rephrasing step, allowing to estimate the marginal contribution of the translator model.

\subsection{Translator}
\label{subsec:translator}
In the previous step we have generated ads from the web pages. We note that this task is much harder than the task of ads rephrasing, with the goal of CTR maximization. Therefore, to improve the generated ads, in the next step we employ a translator model to rephrase the generated ads. 

To this end, we have first trained a translator model over the human-authored ads, used to improve the performance of the ads outputted by the generator model. The proposed ad translation pipeline consists of the following components: (i) Normalization (ii) Preprocessing (iii) Candidate generation (iv) Selection. 

\subsubsection{Normalization}

In order to allow for generalization we created a model to identify medical conditions and proposed treatments and replace them with generic tokens. This is achieved using a custom Named Entity Recognition (NER) model based on the SpaCy library~\cite{spacy}. Every mention of an entity that corresponds to a medical condition or to a treatment was replaced by the generic token $<$CONDITION/TREATMENT$>$.

Specifically, the SpaCy library provides a default NER model which can recognize a wide range of named or numerical entities, including persons, organizations, languages, events, etc. Apart from these default entities, SpaCy also allows the addition of arbitrary classes to the NER model, by training it to recognize new classes. 

The training data for the custom NER consists of sentences, target word(s) in each sentence, and its label. SpaCy also supports the case where a sentence contains more than one entity. For example, for the ad displayed in the first row of Table \ref{tab:ads}, the entities we wish to recognize are both ``Shingles vaccine" and ``shingles" (replacing them with the token $<$CONDITION/TREATMENT$>$). 

The NER was trained by manually labeling $300$ sentences from each domain (see Section \ref{subsec:data} for a definition of the domains), splitting each set for training ($225$ sentences) and test ($75$ sentences). The trained NER model  successfully recognized the entities in $92\%$ and 94$\%$ of the test cases, in each of the two domains. 

%Note that in some of the ads the NER models did not replace correctly the product/service name, which may results in invalid English sentences. 

\subsubsection{Preprocessing}

We used lemmatization and stemming to reduce the vocabulary size  \cite{Korenius:2004:SLC:1031171.1031285}. These two standard techniques transformed inflectional forms and derivationally-related forms of a word to a common base form. We also replaced entity names with their types as follows:
\begin{itemize}
\item A mention of a person name was replaced with the token $<$PERSON$>$. 
\item A mention of a geographic location, such as ``U.S.'' was replaced with $<$GPE$>$. 
\item Organization names (e.g., ``Herbalife'') were replaced with $<$ORG$>$. 
\item Absolute or relative time units, such as ``24/7'', were replaced with $<$DATE$>$. 
\item Monetary values, such as ``$5\$$'' were replaced with $<$MONEY$>$.
\item Numbers were replaced with the token $<$CARDINAL$>$.
\end{itemize}

We used the SpaCy library \cite{spacy} to perform this task. Examples of this process are shown in Table~\ref{tab:ads}. 

\begin{table*}[!ht]
\centering
\begin{tabular}{ |p{40mm}|p{60mm}|p{60mm}| } 
 \hline
Human-authored ad & After preprocessing & Generated Ad  \\ 

 \hline
  \hline
Singling Out Shingles Vaccine - 13 Health Facts. Check out 13 health facts about shingles on ActiveBeat right now. & single out $<$CONDITION/TREATMENT$>$ -  health fact. check out $<$CARDINAL$>$ health fact about $<$CONDITION/TREATMENT$>$ on $<$ORG$>$ right now.
 & $<$CONDITION/TREATMENT$>$ - everything you need to know. discover $<$CARDINAL$>$ fact on $<$CONDITION/TREATMENT$>$. get expert advice now!\\
  \hline
Best Remedy For Cough - Updated 24/7. Search for best remedy for cough. Browse it Now!&  good remedy for $<$CONDITION/TREATMENT$>$ - update $<$DATE$>$. search for good remedy for $<$CONDITION/TREATMENT$>$. browse it now! & home remedy for $<$CONDITION/TREATMENT$>$ - see top $<$CONDITION/TREATMENT$>$ home remedy. try this $<$CARDINAL$>$ effective $<$CONDITION/TREATMENT$>$ remedy that can help you. read more here!\\
\hline
What Does Dark Urine Mean? - Causes Of Dark Urine - Visit Facty, Stay Healthy. See Causes of Dark Urine Color. Learn About What Causes Different Colors Of Urine. &  what do $<$CONDITION/TREATMENT$>$ mean? - cause of $<$CONDITION/TREATMENT$>$ - visit $<$ORG$>$, stay healthy. see cause of $<$CONDITION/TREATMENT$>$. learn about what cause different color of $<$CONDITION/TREATMENT$>$. & $<$CONDITION/TREATMENT$>$ - sign to never ignore. $<$CARDINAL$>$ common $<$CONDITION/TREATMENT$>$ cause. understand how to avoid $<$CONDITION/TREATMENT$>$ and stay healthy.\\
\hline
\end{tabular}
  \caption{Example (human-authored, preprocessed and generated) ads. }
  \label{tab:ads}
\end{table*}

\subsubsection{Candidate generation}

We trained a sequence-to-sequence translator model that learns how to transcribe an input text to an output text, such that the latter corresponds to an ad with a higher CTR value than the input (see Section~\ref{subsec:metrics}).

\paragraph{Training data.}
The training data was constructed as follows: For every search query $q$ we extracted all pairs of ads $a_{\mbox{low}}, a_{\mbox{high}}$ that were presented on the SERP such that $a_{\mbox{low}}$ generated a lower CTR than $a_{\mbox{high}}$ for the same $q$. For every such pair, we generated an example in the training data where $a_{\mbox{low}}$ is the source text and $a_{\mbox{high}}$ is the target text. Note that this process assumes that ads displayed in response to the same query are likely to promote similar products or services.

\paragraph{Model architecture.} We employed a simple sequence-to-sequence (seq2seq) model using the PyTorch library \cite{pytorch}. 
% (For the double-blind submission, the link to our code was removed). 
Data was first tokenized. Then, we built the vocabularies for the source and target ``languages''. Note that even though both the source and the target sentences are in English, they include different sets of sentences, and hence the frequencies of tokens is different.  

The model contains three parts: The encoder, the decoder and a seq2seq model that encapsulates the encoder and decoder.

For both the encoder and the decoder we used a 2-layer Long Short-Term Memory (LSTM) model. The encoder takes as input text and produces a context vector, while the decoder takes the context vector and produces one word at a time.
The complete seq2seq model receives the input text, uses the encoder to produce the context vector, then uses the decoder to produce an output text. 

The optimizer, which is used to update the parameters in the training loop, was set to be the Adam optimizer, and the loss function was set to be the cross-entropy-loss function, which calculates both the log softmax as well as the negative log-likelihood of the predictions.

\subsubsection{Selection}
\label{subsec:ranker}
The translator model can generate multiple candidates, out of which we would like to select the best translation by ranking them. Unfortunately, while for the input ads the CTR values are known, for the output ads these values are unknown. Therefore, here, given an (human-authored or generated) ad, we consider the first ad generated by the translator as the candidate ad. An interesting direction for future research would be to compare between multiple generated ads and to learn how to select the best candidate. Nonetheless, as we show in Section \ref{sec:results}, in the vast majority of cases, the generated ads have succeeded to better attract users interest than their corresponding human-authored ads. 

We note that the proposed pipeline is not entirely automated, and the generated ads contain tokens that should be substituted by the advertiser (see, for example Table \ref{tab:ads}). However, as mentioned in the Introduction,  our goal is to assist health agencies, not completely replace them. Thus, minor grammatical errors in the generated ads may be manually corrected (e.g., replacing words from their lemma form to the required tense), or revised using prior work on the automatic transformation of text to proper English (e.g., \cite{simplenlg}). General tokens (such as $<$CARDINAL$>$ and $<$CONDITION/TREATMENT$>$), if they did not appear in the human-authored ad (e.g., Table \ref{tab:ads} example $1$), may be replaced with the most common original corresponding tokens in the dataset. For example, the most common number for the token $<$CARDINAL$>$ is $10$. Another advantage of the semi-automated pipeline is that advertisers can ensure that the semantic meaning of the ads are maintained in the generated ads.

\section{Experimental Study}

Recall that our goal it to automatically generate ads that attract users' interest.
To better asses the marginal contribution of each of our proposed two models, we consider the following baselines:\\
\begin{enumerate}

    \item \textbf{Human-Authored}: the original ads (to be described in Section \ref{subsec:data}). 
    \item \textbf{Generator} the generator pipeline (described in Section \ref{subsec:generator}), without the rephrasing step, which receives \textit{the human-authored ads' URLs} as the input, and outputs generated ads.
    \item \textbf{Translator}: the translator model (described in Section \ref{subsec:translator}) which receives \textit{the human-authored ads} as the input, and outputs rephrased ads.
    \item \textbf{Generator+Translator} our full model, which receives as the input the URLs of the web pages, generates ads from them, and then rephrase them to improve performance using the trained translator model. 
\end{enumerate}
As we shell see, the ads produced by the Generator baseline behave similarly to the human-authored ads, while the ads produced by the Generator+Translator model are expected to have higher performance. Not surprisingly, as translating a given ad is a simpler task than generating an ad from scratch, we will show that the ads produced solely by the Translator model are expected to get the best performance among all examined baselines. Nonetheless, this model receives the human-authored ads as the input, and thus using solely this model requires health authorities to first write ads by themselves, which, as mentioned in the Introduction, is a task requiring expertise and experience.   

In the following sections we present the experimental setup. 

\subsection{Offline and online metrics}
\label{subsec:metrics}
User interest can be measured in a variety of ways, including clicks, conversions, and future behaviors. In online advertising, the click-through rate (CTR) is the percentage of users viewing a Search Engine Results Page (SERP) who click on a specific ad that appears on that page. CTR measures how successful an ad has been in capturing users' interest. The higher the click-through rate, the more successful the ad has been in generating interest. Moreover, clicks are the most commonly-used measure of ad performance. 

While the goal of many health campaigns is to induce long-term behavioral change, optimizing for this goal introduces challenges such as a delayed and sparse rewards. Therefore, in this work we used CTR as the main success metric. 

\subsection{Data\label{subsec:data}}
%\label{subsec:data}
We extracted English-language search advertisements displayed by customers of Microsoft Advertising between January 1st, 2019 and March 31st, 2019, which were shown at least 100 times. Data is comprised of over $114$K advertisements, displayed to users who queried in two domains: Medical Symptoms (MS) and Preventive Healthcare (PH) (see Table~\ref{tab:data}). 

\begin{table}[!ht]
\centering
\begin{tabular}{ |p{10mm}|p{15mm}|p{15mm}|p{30mm}| } 
 \hline
Domain & $\#$ of advertisements & $\#$ of search queries &  Example keywords \\ 

 \hline
  \hline
  MS & 46061 &  2788 & nasal, vertigo, fatigue, earache, cough.\\
  \hline
    PH & 68021 &  6091 & weight loss, stop smoking, vaccine, safe sex.\\
  \hline

\end{tabular}
  \caption{Extracted Data. }
  \label{tab:data}
\end{table}

To identify ads relevant to these domains, we considered search queries that contain specific (pre-defined) keywords. Keywords for the MS domain comprised of common medical symptoms (according to Wikipedia \cite{symptoms}). Keywords for the PH domain were extracted from a US government website \cite{ph}. In the MS domain, for example, we considered all queries containing the word ``vertigo" (See Table \ref{tab:data} for more examples). We excluded search queries with fewer than $5$ search ads during the data period. The extracted search ads were displayed for over $8$K unique search queries. On average, $14$ different search ads were displayed for each search query.

For the translator, we trained a model separately for each of the two domains. Separate models were used because of the observation that the vocabulary of ads within a domain was more similar than between domains. 
For example, over $35\%$ of ads displayed for the first domain contain one of the words ``remedy'', ``symptom'', ``treatment'', or ``pain'', while over $27\%$ of the ads of the other domain contain one of the words ``help'', ``advice'', ``control'', or ``tip''. We also examined the results while training a single model for both domains, however, the results were inferior, and thus omitted from presentation (see Section \ref{subsec:metrics} for an explanation on how quality was estimated). 

For the generator model, we have trained a single model for both domains. Recall the for the generator model, for each web page, we consider only the corresponding ad achieving the highest CTR. Thus, splitting the data here as well for the two domains result in not enough data for efficient training. In the last step of the generator model, i.e., when using the trained translator model, for each generated ad from the MS (resp., PH) domain we have used the corresponding translator model trained over the data of the MS (resp., PH) domain.

Search ads in Microsoft Advertising include the following elements: (1) A title, which typically incorporates relevant, attention-grabbing keywords; (2)
A URL, to provide users with an idea of where they will be taken once they click on the ad; (3) A descriptive text that highlights the most important details about the product or service, and is used to persuade users to click on the ad. See, for example, the two search ads displayed for the search query ``Shingles vaccine", shown in Figure \ref{fig:serp}. 
In our work we only considered ads' textual components, i.e., the title and the description, concatenated to form a single paragraph. 
Both the title and description fields are limited to a maximum character length. Additionally, both fields have several sub-fields, that is, there are up to $3$ title fields and $2$ description fields, where text needs to be entered into at least one of the title and one of the description fields.

\subsection{Offline evaluation}

Predicting the exact CTR value of a given ad is challenging (as mentioned in Section \ref{sec:related}).
%both linear and non-linear regression models preformed poorly on the task of predicting the CTR of an ad, based solely on its text (as is the case of the generated ad). For example, \textcolor{red}{add details here}.
However, in our setting, it is sufficient to be able to predict the relative order of CTRs for ads on a given query. This is so that the model can predict whether a generated ad is more likely to achieve a higher CTR value than its human-authored version. 
To this end, we employed a learning-to-rank model (LambdaMART~\cite{burges2010ranknet} as implemented by the pyltr library~\cite{pyltr}).

The training data consist of the text of the human-authored ads. To quantify the text we represented it in features such as the sentiment, lexical diversity and the readability index of the text. 
These features were chosen so as to quantify the ability of users to understand and decode the content of ads. 
Previous work has shown the importance of writing a simple, concise, consistent and easily understood text to increase advertising effectiveness \cite{grewal2013quality} \cite{chebat2003testing}. For example, we consider the readability ease of the ads, as measured by the Flesch-Kincaid readability score. 
Other features such as token counts and word embedding were considered, but were found not to improve ranker performance and were thus excluded in the following analysis.
Table \ref{tbl:fetures_ranker} provides the full list of features.

\begin{table*}[!ht]
\centering
\begin{tabular}{ |p{70mm}|p{68mm}|p{26mm}| } 
 \hline
{\bf Feature} & {\bf Explanation} & {\bf Extractor}  \\ 

 \hline
  \hline
Flesch-Kincaid readability ease& Indicating how difficult a sentence in English is to understand. & Textstat \cite{textstat}\\
\hline
Flesch-Kincaid readability grade & Indicating 
the number of years of education generally required to understand the text.& Textstat \cite{textstat}\\
\hline
\# of ``difficult" words & According to the Textstat library \cite{textstat}.& Textstat \cite{textstat}\\
\hline
Readability consensus based upon the Dale-Chall Readability Score, the Linsear Write Formula, and the The Coleman-Liau Index & Indicating the estimated school grade level required to understand the text.& Textstat \cite{textstat}\\
\hline
Vader-Sentiment &Indicating how positive/negative is the sentiment according to the Vader measure.& VaderSentiment  \cite{vader}\\
\hline
Lexical diversity &Number of distinct words divided by the number of words in the text.&\\
\hline
\# of punctuation marks&& SpaCy \cite{spacy}\\

\hline
\# of noun phrases&& SpaCy \cite{spacy}\\

\hline
\# of adjectives&& SpaCy \cite{spacy}\\

\hline
\end{tabular}
  \caption{Features extracted from text ads for the CTR ranker model. }
  \label{tbl:fetures_ranker}
  % \vspace{-8mm}
\end{table*}

The training data consists of lists of extracted (human-authored) ads, along with a partial order between ads in each list. The latter is induced by their CTR values. Each list corresponds to ads shown in response to a different search query. 

The CTR ranker model is used to quantify the effectiveness of the translator and generator models in the following manner: For every human-authored ad and its corresponding translated/generated ads we examine how our trained model ranked the three ads.

Specifically, for each ad we report the rank of (1) the human-authored ad, (2) the translated ad (using the translator model), (3) the generated ad (using only the two first steps of the generator model), and (4) the generated + translated ad (using the full generator model).
%\brit{here}

%If, for a given original ad,
%the generated ad was ranked higher by the ranker, we count this as a positive example (the generated ad is better than the original). Otherwise, it is a negative one. 

\subsection{Online evaluation}

For each of the domain, MS and PH, we selected 10 random ads out of the generated ads for which the ranker ranked higher the generated ads compared to the human-authored ad. We chose to select ads that were estimated to have higher rank because this mimics more closely practical scenarios, where new ads are likely to be used if they are predicted to have superior performance to that of existing ads.

We tested in a real-world setting the performance of the human-authored advertisement compared to the translated ads, the generated ads, and the generated ads after applying translation to them. Note that ads were formatted to fit text length requirements of the ads system (maximum length of the Title and Description fields) by fitting the first sentence of the generated ad into the title field and if it were too long, moving any excess text into the secondary or tertiary title fields. Similarly, the Description field was set. Any minor grammar errors in the generated ads were manually corrected. 

The ads were run as a new advertising campaign (without any history that could bias the experiment) on the Microsoft Advertising network. Each group of ads (i.e., the human-authored and the three algorithmically generated ads) were placed into the same ad group. The campaign was set for automatic bidding with the goal of maximizing CTR, with a maximum bid of US\$$1$ per click. The ads were shown until at least 300 impressions per ad were obtained.

\subsection{Content extraction evaluation}
\label{subsec:content_extraction}

We next examined the Arc90 Readability algorithm's performance, used to extract the main content of the web pages. To this end, we have considered $50$ random web pages. 
For each of these web pages, we have asked $5$ crowdsourced workers from Amazon Mechanical Turk (MTurk, \url{https://www.mturk.com/}) to manually point on the two most relevant paragraphs of the page, which describe best its content. Using the plurality vote aggregation function, two paragraphs for each web page were selected as the most informative paragraphs. Then, we have considered the paragraphs selected by the Arc90 Readability algorithm and compared them with the paragraphs selected by the crowd workers. We report here the number of hits (where the paragraphs selected by the Arc90 Readability algorithm were also selected by the crowd workers) and misses (otherwise).

\subsection{Emotion analysis}
\label{sec:exp_emotions}

We examined three main forms of emotional affect in ads, as described in Section~\ref{subsec:emotion_analysis_res}, namely, the Call-to-Action, arousal and valence, and thought- and feeling-based effects. For each of these affects we measured their value in human-authored and generated ads, and showed the change in their values in each of these ads. 

The Call-to-Action (CTA) of an advertisement is that part of the ad which encourages users to do something~\cite{rettie2005text}. CTAs can drive a variety of different actions depending on the content's goal. CTAs are essential in directing a user to the next step of the sales funnel or process. Previous work has shown that an ad used just to
support a product/service, without a CTA,
might be less effective~\cite{rettie2005text}. 
An ad may contain more than one CTA. For example, in the ad: ``Dry Cough relief - More information provided. Browse available information about dry cough relief. Check here." The CTA are ``browse available information" and ``check here". Here we focus on the verbs of the CTAs (in this example, ``browse" and ``check"), as they are easier to automatically identity (using, e.g., a part-of-speech tagging model).

To classify the emotional attributes of the ad we focused on two concepts which have been shown useful to measure emotional experiences \cite{lang1995effects} \cite{kensinger2004remembering}: {\bf Arousal} and {\bf Valence}. Arousal refers to the intensity of an emotion (how calming or exciting it is) and valence deals with the positive or negative character of the emotion. An ad with positive connotations (such as joy, love, or pride) is said to have high valence. Negative connotations (including death, anger, and violence) have low valence. Similarly, the more exciting, inspiring, or infuriating an ad is, the higher the arousal. Information that is soothing or calming produces low arousal. 

To quantify CTA, arousal and valence we first created a dataset of ads labeled for their arousal and valence levels, and well as marking the CTA verbs therein. Then, a model for each was created using this dataset, and applied to a larger set of ads. 

The dataset was created from a random sample of $265$ advertisements, comprising of generated and human-authored ads, as follows: $83$ human-authored MS-related ads, $82$ generated MS-related ads, $50$ human-authored PH-related ads, and $50$ generated PH-related ads.

The values of valence, arousal and the CTA verbs for the dataset were found by asking $5$ crowdsourced workers from Amazon Mechanical Turk  (MTurk, \url{https://www.mturk.com/}) to label the $265$ ads. Specifically, workers were asked to mark CTA verbs and to estimate (separately) the arousal and valence scores of each ad on a $5$-point scale in the range of $[-2,2]$, where $-2$ is the lowest arousal/valence score and $2$ is the highest score. A score of $0$ suggests that the ad is neutral with respect to arousal/valence experience.

Using the training set we created three models: one to identify CTA verbs, and another two to estimate valence and arousal in non-labeled ads. 

CTA verbs in other (non-labeled) ads were identified by implementing a custom Part-of-Speech (PoS) tagger using the SpaCy library \cite{spacy}. The trained model was applied to non-labeled ads, to tag all words. Every word that was tagged as CTA was counted as a CTA verb.

As for the valence and arousal scores, two models were trained: One to predict the average arousal score reported by MTurk workers and the other to predict the average valence score. The features of these models were set to be the tokenized text of the ads, using the ML.NET tool (\url{https://dotnet.microsoft.com/apps/machinelearning-ai/ml-dotnet}).

All models were constructed using the Microsoft ML.NET machine learning tool. We examined multiple models, including linear regression (with stochastic gradient decent training), boosted trees and random forest. The best results were achieved with the Boosted Trees and Random Forest regression models, for the arousal and valence scores, respectively. In the Boosted Trees model the number of trees was set to $100$ and the learning rate was set to $0.2$. In the Random Forest regression model the number of trees was set to $500$. The performance of the models on training data was evaluated
using $5$-fold cross-validation.

\begin{table*}[!ht]

\begin{center}
\begin{tabular}{ |p{20mm}|p{25mm}|p{35mm}|p{78mm}| } 
    \hline
    {\bf Effect}   & {\bf User's Desire}    & {\bf Examined key words}& {\bf Example ad}   \\
 \hline
 \hline
   Thought-based &Petty advantage& discount, deal, coupon, $x\%$&``Science diet coupons - Up to 60\% Off Now. Christmas Sales! Compare..."\\
    \hline
   Thought-based & Extra convenience& delivery, payment, shipping& ``Unbearable Smokeless Coals - Great Range, Fast Delivery..."\\
    \hline
    Feeling-based&Trustworthy& official, guarantee, return& ``Jenny Craig Official Site - A Proven Plan For Weigh Loss..." \\
    \hline
     Feeling-based&Luxury seeking&top, most, best, good&`` Best Remedy For Cough - Updated 24/7..."\\
    \hline
\end{tabular}
\end{center}
\caption{Examined user desires under thought-based and
feeling-based effects. }
\label{tab:effects}
\end{table*}

To examine the thought-based and feeling-based effects of the ads, we considered user desires under both effects, as proposed in Wang et al. \cite{wang2013psychological}. We used the phrases proposed in that paper, which were extracted from general ads by human experts, adapting them to those which we found appeared frequently in ads related to the MS and PH domains. Table \ref{tab:effects} lists these effects, their associated user desires, and the keywords that were used to identify these effects in the ads. Wang et al.  \cite{wang2013psychological} used a novel algorithm to mine these user desire patterns. Here we used predefined keywords to conclude if a given ad encompasses one of the examined user's desires. 

In Section \ref{subsec:emotion_analysis_res} we discuss how the proposed translator model has learned to take advantage of these user desires, to increase the likelihood a user will click on an ad (i.e., to increase users' interest).

\section{Results}
\label{sec:results}
In this section we provide results of our efforts to validate the effectiveness of the proposed pipelines. We also aim to explain what each of the models has learned.

\subsection{Content Extraction Evaluation}
As mentioned in Section \ref{subsec:content_extraction}, we evaluated the performance of the Arc90 Readability algorithm via a user study. We report that in the vast majority of the examined cases, the algorithm and the participants have selected the same paragraphs. In particular, in $84\%$ of the cases, the algorithm and the participants have chosen the same paragraph as the most important one, and in $81\%$ of the cases, the algorithm and the participants have chosen the same paragraph to be the second most important paragraph. Our results show that identifying the most relevant paragraph is easier than defining the second most relevant paragraph. Indeed, in 78\% of the cases, three or more (out of 5) crowd workers have chosen the same paragraph to be the most informative paragraph for a given web page. In comparison, only in $56\%$ of the cases three or more crowd workers have chosen the same paragraph to be the second-best paragraph. This implies that
even in the cases where the second paragraph selected by the Arc90 Readability algorithm was counted as ``miss", it may nevertheless be relevant.

% \begin{table*}[h]

% \begin{center}
% \begin{tabular}{ |p{28mm}|p{25mm}|p{25mm}| } 
%     \hline
%     {\bf Paragraph}   & {\bf Hits}    & {\bf Misses}  \\
%  \hline
%  \hline
% First Paragraph &84\%&16\%\\
%     \hline
%     Second Paragraph &81\%&19\%\\
%     \hline
% \end{tabular}
% \end{center}
% \vspace{-3mm}
% \caption{Performance evaluation of the Arc90 Readability algorithm via a user study. }
% \label{tab:acrc90}
% \end{table*}

\subsection{Offline estimation of result quality}
\label{res:offline}

\begin{figure}[!ht]
\centering
\includegraphics[width=0.8\linewidth]{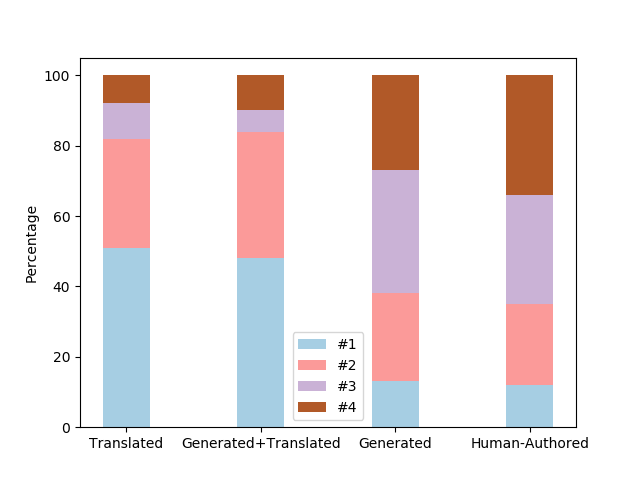}
\vspace{-10px}
\caption{Percentage of ads of each baseline, to be placed at each rank (1-4), according to the CTR ranker. }
\label{fig:ctr_offline}
\end{figure}

Using the trained CTR ranker, for each human-authored ad, we examine the ranks of its corresponding generated ads (produced by our proposed models). For each (human-authored or auto-generated) ad the CTR ranker outputs probabilities reflecting its estimated rank. Ads in which their probabilities difference was less than $0.1$ were treated as having the same rank. 
According to the Kemeny–Young method~\cite{KemenyYoung}, the optimal ranking is: 1. Translator 2. Translator+Generator 3. Generator 4. Human-authored. 
The full results are depicted in Figure \ref{fig:ctr_offline}.

Overall, the CTR ranker predicts that the ads generated by the translator model will have the highest CTR in 51\% of the cases. This result is not surprising, since, as mentioned, translating a given ad to improve its performance is a much simpler task than generating an ad from a web page. The second-best competitor is the full generator model (i.e., Generated+Translated), where the CTR ranker predicts that the ads generated by this model will have the highest CTR in 48\% of the cases (in 38\% of the cases, the differences between the probabilities were less than $0.1$).  

Observe that adding the rephrasing step to the generator model significantly improves its performance. Namely, the ads generated only by the generator model (without the last rephrasing step) are predicted to have the highest rank only in $13\%$ of the cases. As can be seen, our results indicate that ads produced by the generator model behave similarly to the human-authored ads. This result is surprising considering the difficulty of the task of generating an ad directly from a given web page.

We note, however, that the CTR ranker cannot be considered as a perfect proxy for experimentation, since it does not precisely predict the ordering of ad pairs by their performance. To estimate the likelihood of ranker error (where it will rank higher the ad with the lower CTR in a pair) we report the Kendall's Tau (KT) rank correlation of the CTR ranker. This measure was chosen as it considers the number of concordant and discordant pairs of the prediction, compared to true performance.  
The CTR ranker model was evaluated
using 5-fold cross validation (over the human-authored ads). The average KT across folds was $0.44$ ($P < 0.01$). This should be compared to KT equalling $0.36$ when randomly ordering the ads. Thus, because the CTR ranker is imperfect, some of the generated ads which are predicted by the ranker to be less effective than their corresponding human-authored ads may achieve higher CTR values in practice, and vice versa.

\subsection{Online estimation of result quality}
Twenty ads were run on the Microsoft Advertising platform between March 3rd and April 28th, 2020, until they were shown at least 300 times each. The average number of times that each ad was shown was 1,74 times, and the maximum was 6,191 times.  

\begin{figure}[!ht]
\centering
\includegraphics[width=0.8\linewidth]{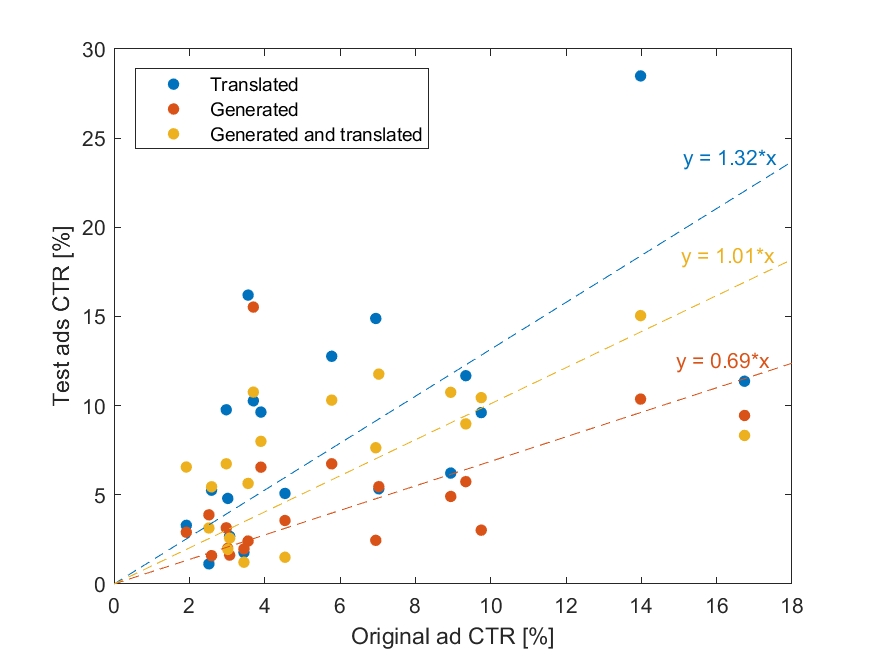}
\vspace{-10px}
\caption{CTR of the generated ads versus the human-authored ads. Each dot represents one advertisement pair (one human-authored versus one generated by each of the three models). }
\label{fig:ctr}
 \vspace{-4mm}
\end{figure}

Figure \ref{fig:ctr} shows the CTR of the generated ads versus those of the human-authored ads, for each type of generated ad. As the figure shows, the generated ads received, on average, approximately 28\% fewer clicks than the human-authored ads ($P=0.046$, signrank test). The translated ads received 32\% more clicks than the human-authored ads ($P=0.040$, signrank test), and the translation of the generated ads had approximately the same CTR as the human-authored ads (not statistically significant). 

These results indicate that our proposed framework, i.e., the Generator+Translator model, can generate entirely new ads, which their performance are highly similar to the ads that were written by experts. Moreover, the results show that the translator model can significantly improve the performance of the current health-related ads, by simply rephrasing them.  
% In 15 of 19 cases generated ads had a higher CTR than the original ads. On average, the generated ads had a CTR which was 68.2\% higher than that of the original ads (statistically significant, signrank, $P=0.011$). 

% The average per-advertisement improvement in CTR (micro-averaged improvement) was 25.4\%. Note, however, that this average refers to only 15 ads, since, as Figure \ref{fig:ctr} shows, 4 of the ads had a non-zero CTR for the generated version, but zero for the original version.

% \begin{figure}
%     %\centering
%     \includegraphics[scale = 0.8]{figures/Summary.png}
%     \vspace{-10px}
%     \caption{Performance evaluation of the Arc90 Readability algorithm.}
%     \label{fig:acrc90}
% \end{figure}

\subsection{Emotion analysis results}
\label{subsec:emotion_analysis_res}

To obtain an understanding for what the proposed models have learned, we examined the frequency of using Call-To-Actions (CTAs) and the emotional attributes of the generated ads.

\subsubsection{CTAs analysis}

Previous work has shown that users tend to click on ads that can
convince them to take further actions, and the critical factor is if
those ads can trigger users' desires \cite{wang2013psychological}. 
Furthermore, Rettie et al. \cite{rettie2005text} demonstrated a relationship between the level of relevance and action taken, such that when users
found ads relevant they were
significantly more likely to take action (i.e., to click on the ads). Thus, an ad used solely to inform of a product or service, without containing a CTA,
might be less effective.

We used the custom PoS tagger (see Section \ref{sec:exp_emotions}) to identify CTA verbs in both human-authored and generated ads. The performance of the model was evaluated
using 5-fold cross validation on the labeled ads (i.e., the ads that were labeled by the crowd workers), and we report
the average score across all runs. On average, the PoS tagger correctly tagged $93\%$ of the CTA verbs.

\begin{figure}[!ht]
\centering
\includegraphics[width=0.8\linewidth]{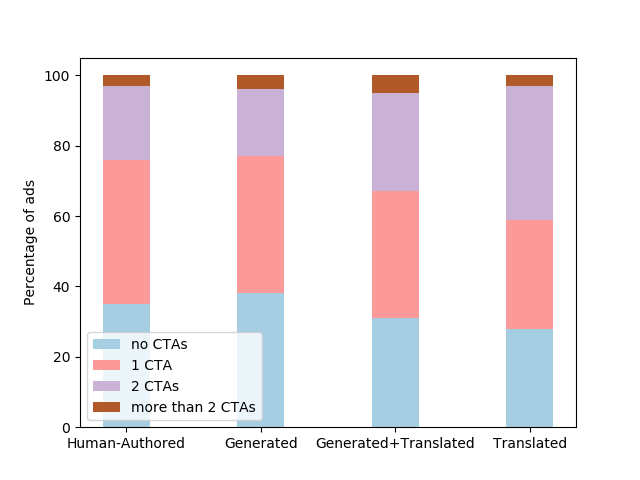}
\vspace{-10px}
\caption{Predicted number of CTAs.}
\label{fig:cta}
\end{figure}

Analysis reveals that $72\%$ of the ads generated by the translator model include at least one CTA verb, compared with only $65\%$ of the human-authored ads including at least one CTA verb (statistically significant, signrank test, $P<0.05$). Regarding the generator model, without the last rephrasing step, only $62\%$ of the ads generated by this model include at least one CTA verb, compared with $69\%$ of the ads generated by the full generator model including at least one CTA verb (statistically significant, signrank test, $P<0.05$). 
According to the signrank test, the human-authored ads and the ones produced by the generator model were found to be not statistically significant. Similarly, the ads produced by the full generator model (i.e., generator+translator) and the translator were found to be not statistically significant. Nonetheless, we report that all of the results of the baselines between these two groups are indeed statistically significant (signrank test, $P<0.06$).  
%\brit{which statistic test should I use here?}
%(statistically significant, chi-square test, $P<0.05$).  
% The most frequently used CTA verbs in the original and in the generated ads are shown in Table \ref{tab:ctas}.
% We note that in the MS domain, where users are seeking  information on a medical symptom, a desired result contains information on the symptom and ways to treat it. Examining the table, the two most common CTA verbs in the generated ads in the MS domain are ``learn" and ``treat". Namely, the translator has learned that these CTAs are very effective in triggering users' desires, and therefore the use of them leads to an increase of the CTR. Similarly, in the PH domain, where, for example, users seeking information on ways to quit smoking or to lose weight, the translator often used CTAs which are relevant to this goal (e.g., ``quit", ``stop", ``lose"). 

% \begin{table}[h!]
% \centering
% \begin{tabular}{ |p{10mm}|p{30mm}|p{30mm}| } 
%  \hline
% Domain & Original ads & Generated ads \\ 

%  \hline
%   \hline
%   MS & read, learn, treat, browse, check. &  learn, treat, find, discover, get. \\
%   \hline
%     PH & get, learn, visit, read, quit. &  quit, get, stop, learn, lose.  \\
%   \hline

% \end{tabular}
%   \caption{Top $5$ CTA verbs in the original and generated ads. }
%   \label{tab:ctas}
%   \vspace{-10mm}
% \end{table}

\subsubsection{Arousal and valence analysis}

\begin{figure}[!ht]
\centering
\includegraphics[width=0.8\linewidth]{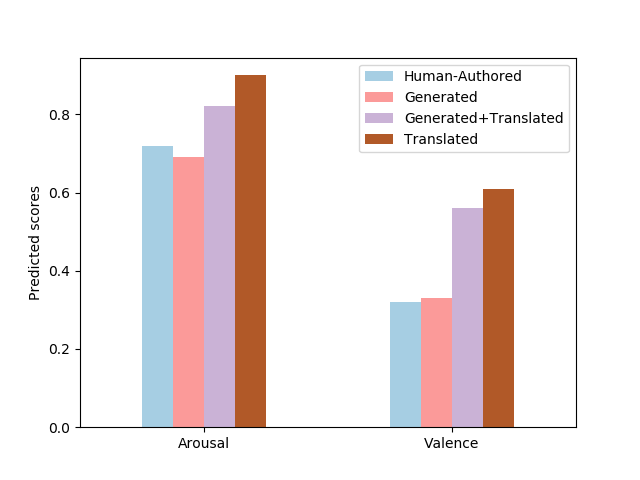}
\vspace{-10px}
\caption{Predicted arousal and valence scores.}
\label{fig:arousal_valence}
\end{figure}

To asses how well the crowd workers agree among themselves in the arousal and valence of ads, we computed the average (valence/arousal) score for each advertisement and compare the standard deviation of the scores around this average value. These are compared with the standard deviation of random scores. A lower standard deviation for crowd worker scores (compared to random) indicates greater agreement among workers.

The standard deviation of the arousal scores was $0.23$ (compared to $0.78$ for random), and 
the standard deviation of the valence scores was $0.19$ ($0.75$ for random scores). Thus, workers were able to agree on the valence and arousal scores of ads to a very high degree. 

Using $5$-fold cross-validation on the crowdsourced-labeled data, the models predicted arousal and valence from ad text with an accuracy of $R^2=0.62$ and $0.42$, respectively. 
The trained models were then applied to predict the arousal and valence scores of all human-authored and generated ads. 

The averaged predicted arousal and valence scores of the human-authored and of the generated ads are depicted in Figure \ref{fig:arousal_valence}. Note that even though these values are measured on a scale of $[-2,2]$, most of the predicted scores were positive, and so are the average scores.

Interestingly, observe that the predicted arousal and valence scores of the ads generated by the translator model are the highest. This implies that the proposed translator model is predicted to increase the arousal and valence of the input ads (i.e., the human-authored ads or the ads generated by the generator model).

\begin{figure*}[!ht] 
  \subfigure[Arousal - Translated]{% 
    \includegraphics[width=.32\textwidth]{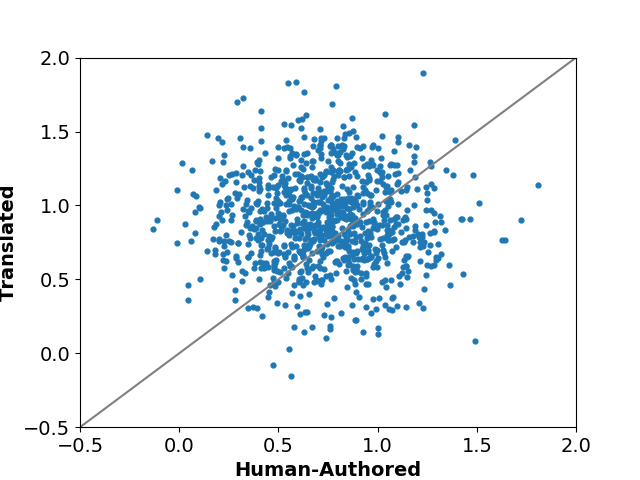}  
  } 
  \subfigure[Arousal - Generated + Translated]{% 
    \includegraphics[width=.32\textwidth]{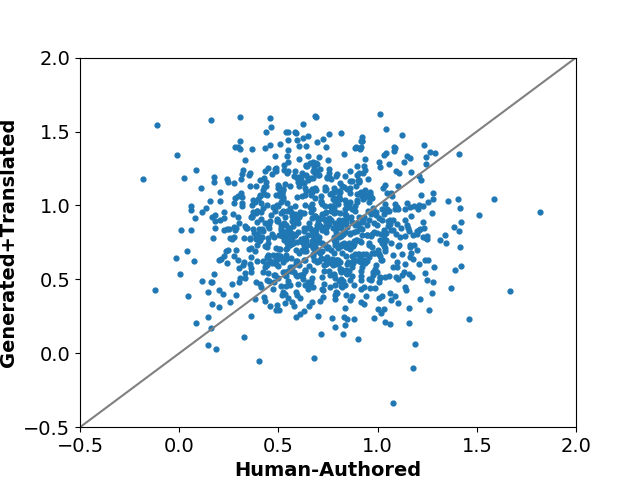} 
  } 
    \subfigure[Arousal - Generated]{% 
    \includegraphics[width=.32\textwidth]{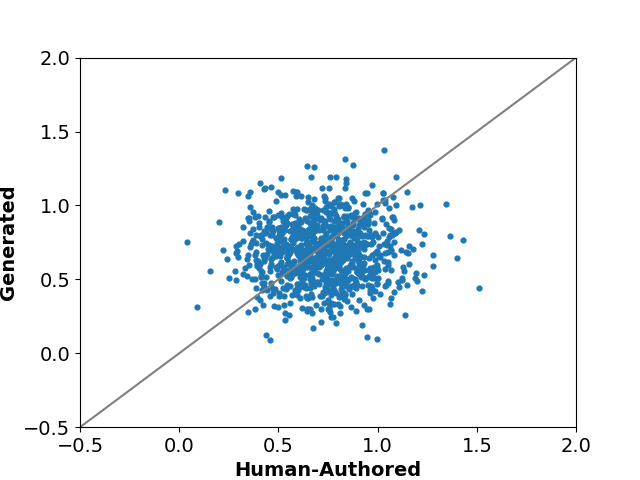}  
  }
  
  ~~
   \subfigure[Valence - Translated]{% 
    \includegraphics[width=.31\textwidth]{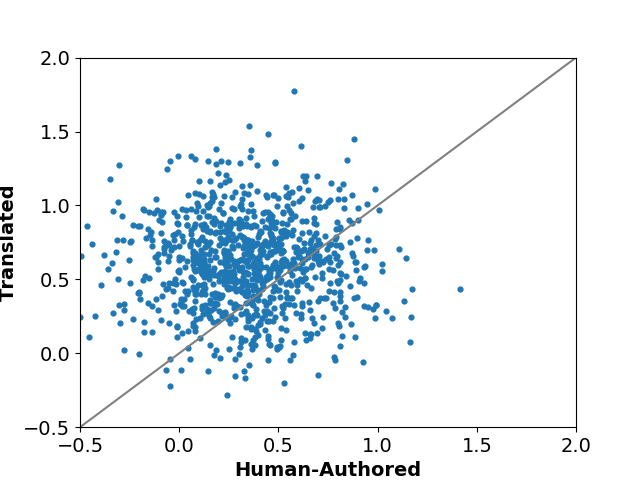}  
  } 
  \subfigure[Valence - Generated + Translated]{% 
    \includegraphics[width=.31\textwidth]{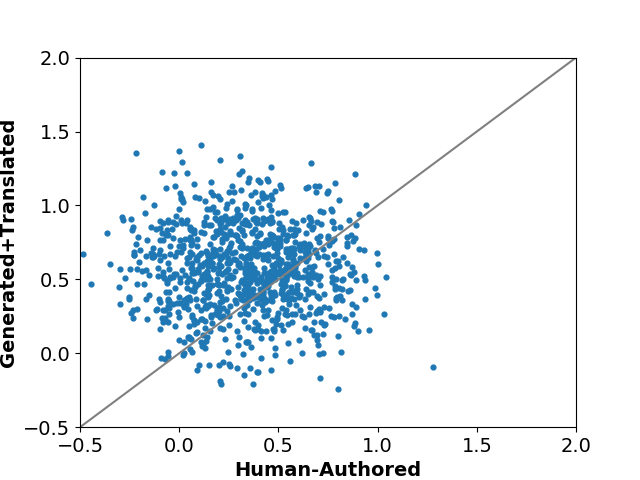}  
  } 
    \subfigure[Valence - Generated]{% 
    \includegraphics[width=.31\textwidth]{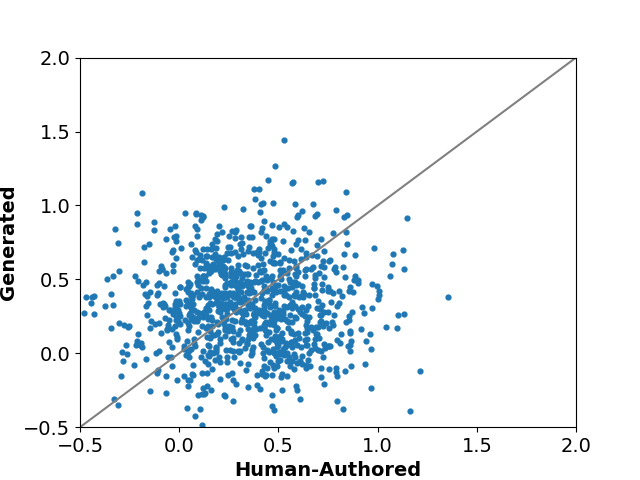} 
  }
 \vspace{-4mm}
  \caption{Predicted arousal/valence scores of $1$k random human-authored ads vs. those of the ads generated by the baselines. The diagonal line indicates equal values. }
  \vspace{-4mm}
  \label{fig:arousal_valence_scores1}
\end{figure*}

Figure \ref{fig:arousal_valence_scores1} shows the predicted valence and arousal scores of 1000 random human-authored ads versus those of the ads generated by the proposed models. The diagonal line indicates equal values for the human-authored and auto-generated ads. Points on or below the grey line denote ads for which the auto-generated ads had equal or lower valence or arousal scores, compared to those of the human-authored ads. As the figures show, for the translator and the full generator (i.e., generated + translated) models, for both arousal and valence the vast majority of points (i.e., ads) are above the grey lines. Thus, in most cases, the ads generated either by the translator or by the full generator models are predicted to have higher arousal and valence scores than their corresponding human-authored ads, according to our models. In contrast, as can be seen, only slightly more than half of the ads generated only by the generator model are predicted to have higher arousal or valence scores than their corresponding human-authored ads. 

\begin{figure}[t]
\centering
\includegraphics[width=0.8\linewidth]{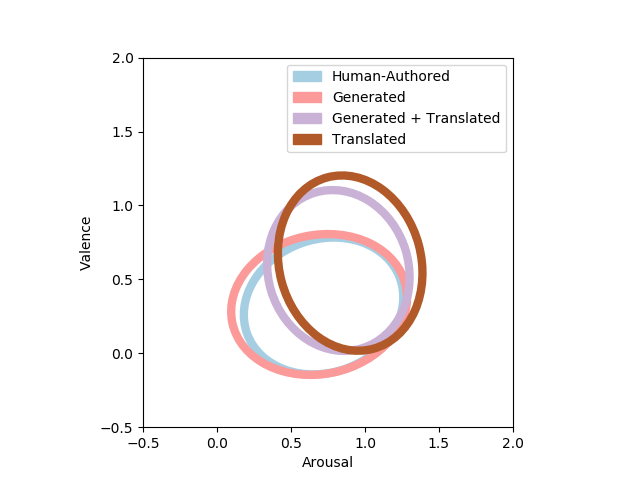}
\vspace{-10px}
\caption{Predicted valence versus arousal scores of $1$k random human-authored ads and their corresponding auto-generated ads. Ellipses denote one standard deviation of the mean.}
\label{fig:arousal_valence_scores2}
\end{figure}

Figure \ref{fig:arousal_valence_scores2} depicts the predicted valence scores versus the predicted arousal scores of (the same) 1000 random human-authored ads and their corresponding auto-generated ads of the three models. Ellipses denote one standard deviation around the mean. As can be seen, on average the ads generated by either the translator or by the full generator model are predicted to have higher arousal and higher valence, compared to the human-authored ads. That is, these models model increase both the arousal and the valence scores of the ads, compared with their human-authored versions. In contrast, the ads produced by the generator model are rather similar to the human-authored ads and there is almost no change in the predicted arousal and valence scores.   

Last, we examine the correlation between users' interest (measured, in our setting, by the CTR values) and the predicted values of the arousal and valence. Specifically, we report the Pearson correlation between CTR and these values. We note that, as mentioned in Section \ref{sec:related}, as opposed to arousal, where it has been shown that ads with high arousal increase user attention, ads with
very positive (i.e., a score close to $2$) or very negative (i.e., a score close to $-2$) valence are similarly likely to increase users' interest, compared to  ads with neutral valence scores (i.e., a score close to $0$). Therefore here we consider the absolute valence scores. The Pearson correlation between the CTR of the human-authored ads and the predicted arousal score was $0.72$ ( $P<0.01$), and the Pearson correlation between the CTR of the human-authored ads and the absolute value of the valence score was $0.62$ ( $P<0.01$). Namely, we observe a moderate positive relationship between CTR and the arousal and (absolute) valence scores. Thus, as the arousal increases it is more likely that users would click on the ads. Similarly, as the absolute valence score increases so is the chance a user would click on the ad.

\subsubsection{Analysis of thought- and feeling-based effects}

As mentioned in Section \ref{sec:related}, the inclusion of specific textual content referring to user desires increases user interest, and consequently CTR \cite{wang2013psychological}. To investigate if these user desires are associated with an increased CTRs, we computed the CTR of (human-authored) ads containing the keywords associated with each desire and compared them with the average CTR of all ads that were displayed to the same search query. The results are shown in Table \ref{tab:effects_ctr_}. 
Indeed, one can see that the likelihood a user will click on an ad increases if it contains one of the keywords mentioned in Table \ref{tab:effects}. 

\begin{table}[!ht]

\begin{center}
\begin{tabular}{ |l|c|c| } 
%\begin{tabular}{ |p{30mm}|p{20mm}|p{20mm}| } 
    \hline
    {\bf Effect}   & {\bf Percentage of}    & {\bf Change in CTR }   \\
                  & {\bf matched ads}      &                     \\
 \hline
 \hline
 Petty advantage &$7.2$&$48.5\%$\\
    \hline
     Extra convenience &$6.8$&$35.7\%$\\
    \hline
    Trustworthy &$2.8$&$28.1\%$\\
    \hline
     Luxury seeking &$4.9$&$33.2\%$\\
    \hline
\end{tabular}
\end{center}
\caption{Change in CTR between ads matched with the patterns of each effect and other ads that were displayed in response to the same query.}
\label{tab:effects_ctr_}
\end{table}

Additionally, we examined the percentage of human-authored and auto-generated ads containing at least one of the keywords associated with each of the examined user desires, as listed in Table \ref{tab:effects}.

\begin{figure}[t]
\centering
\includegraphics[width=0.8\linewidth]{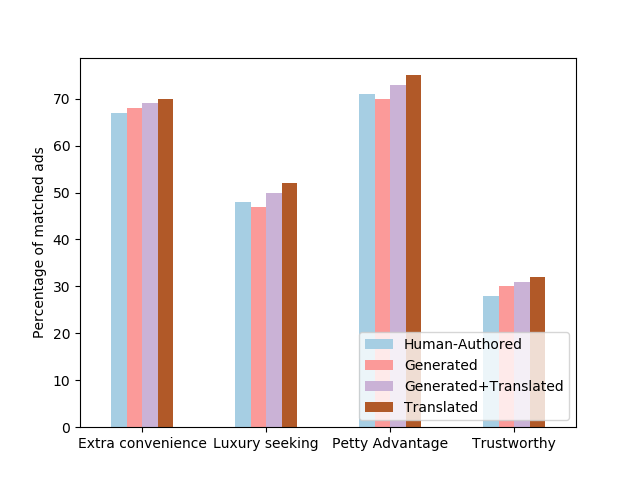}
\vspace{-10px}
\caption{Percentage of ads containing at least one of the keywords associated with each of the examined user's desires.} % Differences for each of the $4$ values is statistically significant (paired signtest, $P < 10^{-2}$). }

\label{fig:effects_res}
\end{figure}

The results are shown in Figure \ref{fig:effects_res}. Here, the vertical axis represents the percentage of ads containing at least one of the keywords associated with each of the examined user desires. Observe that in all cases, the ads generated either by the translator or the generator models include more such keywords compared with the human-authored ads, indicating that both models have learned that incorporating specific textual content increases user interest.  

Thus, our results support the empirical results of Wang et al. \cite{wang2013psychological}, showing that users tend to click more on ads containing specific patterns and keywords.    

\section{Discussion}

In this study we presented a system which given an health-related web page, it generates an optimized ad that is expected to be clicked by users.
An immediate application of this work is to offer such a service to public health authorities, which may improve population health by more effectively eliciting positive behavioral change. 

Beyond training of the models, our work requires several supporting models for evaluating the resulting ads. These include a ranking model to estimate improvement in CTR, and models for assessing psychological attributes of original and generated ads. Moreover, we tested the our proposed pipelines, separately and together, in a real-world scenario by running $20$ advertisement in a head-to-head competition, and found that our auto-generated ads are expected to behave similarly as the original ad. This surprising result indicated that our proposed framework can assists public health authorities to automatically generate ads. In case the end-user (i.e., health authority) provides an input ad, our results indicates that, using solely the translator model, we can improve its performance in $32\%$ (on average).  

% a $68\%$ performance improvement in the generated ads, compared to the original ones.

To investigate what our models have learned, we examined $3$ factors of the original ads compared to those of the automatically generated ads: (1) The use of Calls-to-Action (CTAs), which have been shown to be essential in increasing the ads effectiveness \cite{rettie2005text}; (2) The estimated arousal and valence scores of the ads, where previous work has shown that the inclusion of high arousal and valence sequences in ads increases user attention and interest~\cite{belanche2014influence}; and (3)
The inclusion of special textual content that refer to the desires of users \cite{wang2013psychological}. 

Our empirical results indicate that the translation model improved all $3$ factors. Specifically, the ads generated by this model include more CTAs than the original or auto-generated ads, they are predicted to have higher arousal and valence emotions, and they combine more keywords which have been shown related to user desires. Thus, the translation model has, without explicit guidance, learned which psychological factors should be enhanced in ads so as to elicit higher CTR. 

Our work enables advertisers, especially in the health domain, to create advertising campaigns without advertising expertise. This new area of work includes several future directions for research.

First, here we focused on improving CTR. As was discussed in the Introduction, other measures of ad performance, including conversion and future behavior, may be more useful for health authorities. Training a model to improve these will likely require training on sparser data, but the resulting model may be of better use to health authorities, and comparing the psychological attributes it affects will be of interest to marketing experts.

% Another interesting direction for future research will be to generate ads directly from product or service web pages, to assist health authorities in writing effective ads without using (possibly expensive) experts and experiments. 
Another interesting direction for future research will be to apply our algorithm to more domains, such as wellness and education and build a more general model, that is, one which is suitable for any health-related ad, not to one of a specific domain therein. 

On the technical side, we assume that performance may be improved if a more complex translator model was used (see discussion in the Related Work), and if different outputs of the translator and generator models were examined for the same input ad, tuning the models to achieve better results. Lastly, recall that the translator model receives as input ads in their basic form (i.e., after pre-processing, see Table \ref{tab:ads}), and therefore the generated ads are also in this form. Future work will improve the automatic transformation of the ads to proper English (e.g., realizationing words from lemma form using~\cite{simplenlg}).   

Additionally, it may be possible to enable advertisers to manually tune the preferred psychological values of the output, by intervening within the intermediate layers of the translation model. 

Finally, here we focused on textual ads. Much of the advertising industry uses imagery and audio. It will be interesting to try and improve those through our models. 

%\newpage

%\newpage
\bibliographystyle{ACM-Reference-Format}
\bibliography{main}

\end{document}